\documentclass[11pt,letterpaper]{article}
\pdfoutput=1

\usepackage[margin=1in]{geometry}
\usepackage[section]{algorithm}
\usepackage[noend]{algorithmic}

\usepackage{microtype}

\usepackage{hyperref}

\usepackage{amsfonts}
\usepackage{amssymb}
\usepackage{textcomp}
\usepackage[shortlabels]{enumitem}
\usepackage{graphicx}
\usepackage{subfigure}

\newtheorem{theorem}{Theorem}[section]
\newtheorem{lemma}[theorem]{Lemma}

\newtheorem{corollary}[theorem]{Corollary}

\newtheorem{definition}[theorem]{Definition}

\newboolean{short}
\setboolean{short}{true}

\newcommand{\onlyShort}[1]{\ifthenelse{\boolean{short}}{#1}{}}
\newcommand{\onlyLong}[1]{\ifthenelse{\boolean{short}}{}{#1}}
\newcommand{\shortLong}[2]{\ifthenelse{\boolean{short}}{#2}{#1}}
\newcommand{\longShort}[2]{\ifthenelse{\boolean{short}}{#2}{#1}}

\onlyShort{
  \renewcommand{\paragraph}[1]{\noindent\textbf{#1}}
}

\newcommand{\squishlist}{
 \begin{list}{$\bullet$}
  { \setlength{\itemsep}{0pt}
     \setlength{\parsep}{3pt}
     \setlength{\topsep}{3pt}
     \setlength{\partopsep}{0pt}
     \setlength{\leftmargin}{1.5em}
     \setlength{\labelwidth}{1em}
     \setlength{\labelsep}{0.5em} } }

\newcommand{\squishlisttwo}{
 \begin{list}{$\bullet$}
  { \setlength{\itemsep}{0pt}
     \setlength{\parsep}{0pt}
    \setlength{\topsep}{0pt}
    \setlength{\partopsep}{0pt}
    \setlength{\leftmargin}{2em}
    \setlength{\labelwidth}{1.5em}
    \setlength{\labelsep}{0.5em} } }

\newcommand{\squishend}{
  \end{list}  }
  
  \usepackage{textcomp}

\makeatletter
\DeclareRobustCommand*\cal{\@fontswitch\relax\mathcal}
\makeatother

\title{
On The Termination of a Flooding Process
}

\author{Walter Hussak \thanks{Computer Science,  Loughborough University, UK, \texttt{W.Hussak@lboro.ac.uk}}
\and Amitabh Trehan \thanks{Computer Science,  Loughborough University, UK, \texttt{A.Trehan@lboro.ac.uk}}}

\date{}

\begin{document}

\begin{titlepage}

\maketitle


\begin{abstract}
\normalsize
Flooding is among the simplest and most fundamental of all distributed network algorithms. 
In a synchronous message passing network, for example, node(s) begin the process by sending a message to all their neighbours and the neighbours, in the next round forward the message to all the neighbours they did not receive the message from and so on. We assume for various reasons (simplicity and memory considerations), the nodes do not keep a record of the flooding event. We call this amnesiac flooding. Since the node forgets, if the message is received again in subsequent rounds, it will be forwarded again raising the possibility that the message may be circulated infinitely even on a finite graph.  As far as we know, the question of termination for such a flooding process has 
not been settled -  rather, non-termination is implicitly assumed.

In this paper, we show that synchronous amnesiac flooding always terminates on any arbitrary finite graph and derive exact termination times which differ sharply in bipartite and non-bipartite graphs. Let $G$ be a finite connected graph. We show that synchronous flooding from a single source node terminates on $G$ in $e$ rounds, where $e$ is the eccentricity of the source node, if and only if $G$ is bipartite. For non-bipartite $G$, synchronous flooding from a single source terminates in $j$ rounds where $e < j \leq e+d+1$ and $d$ is the diameter of $G$. 
   Since $e$ is bounded above by $d$, 
   this implies a termination time of 
   at most $d$ and of 
   at most  $2d + 1$ for bipartite and non-bipartite graphs respectively. If communication/broadcast to all nodes is the motivation, our result shows that the simple flooding process is asymptotically time optimal and obviates the need for construction and maintenance of spanning structures like spanning trees. Moreover, the clear separation in the termination times of bipartite and non-bipartite graphs may suggest  possible mechanisms for distributed discovery of the topology/distances in an arbitrary graph.
   
   For comparison, we also show that, for asynchronous networks, however, an adaptive adversary can force the process to be non-terminating. 

\end{abstract}

\end{titlepage}

\section{Introduction}

\begin{figure}[h!]%
\centering 
\subfigure[The Hypercube graph (on 8 nodes)]{ \includegraphics[scale=0.2]{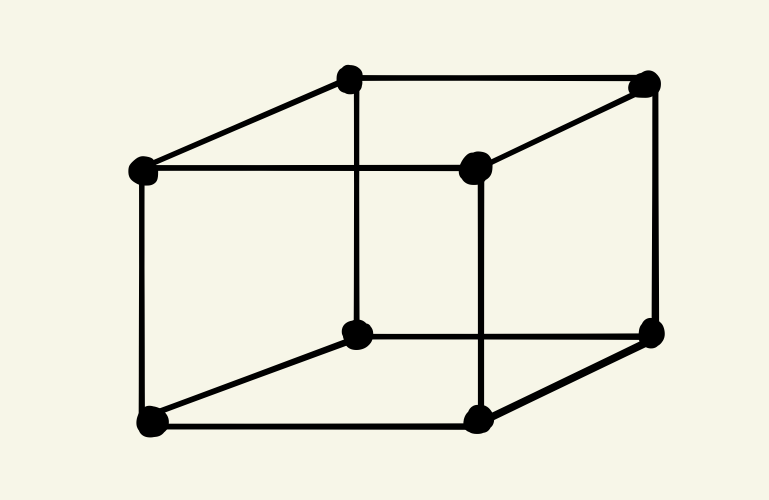}}\label{fig: hcube}
\quad 
\subfigure[The Petersen Graph]{\includegraphics[scale=0.2]{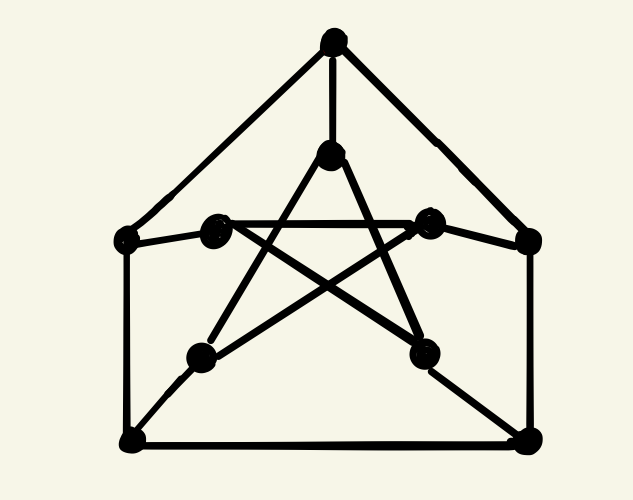}}\label{fig: pgraph} \\
 \subfigure[Flooding on Hypercube]{  \includegraphics[scale=0.23]{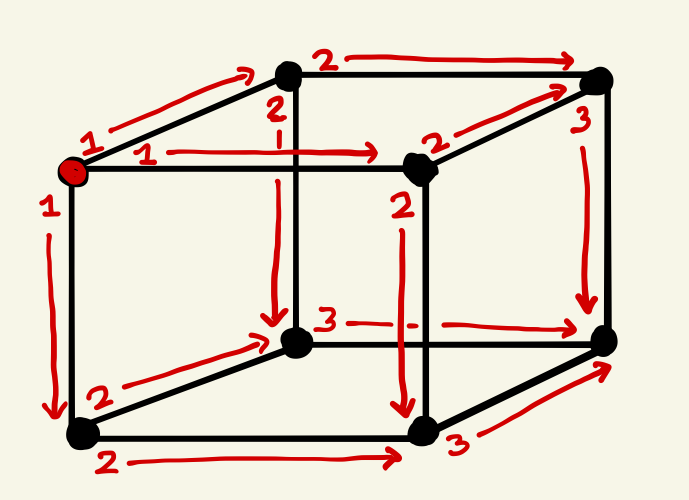}} \label{fig: hcubeflood}
 \quad
 \subfigure[Flooding on the Petersen Graph]{  \includegraphics[scale=0.23]{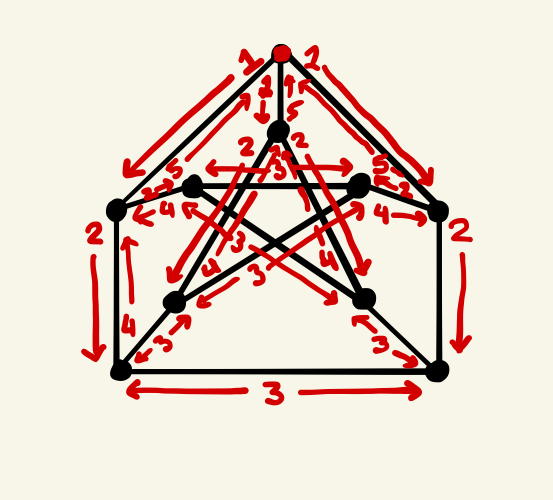}}\label{fig: pgraphflood}
\caption{Two well known graph topologies (Hypercube and the Petersen Graph) and execution of Amnesiac flooding (from the red coloured node) on them. The arrows point to direction of the transmission of the message with the label giving the round number. Double headed arrows indicate the message crossing over in both directions on the edge. The flooding on the hypercube terminates in only  $3$ = diameter rounds, whereas on the Petersen graph, it takes $5$ = 2* diameter + 1 rounds.}
        \label{fig: floodingline}
\end{figure}

Consider the two well known graphs in Figure~\ref{fig: hcube}; the hypercube (cube in 3 dimensions) graph and the Petersen graph. Now, consider distributed networks where nodes follow the following simple flooding process as a communication primitive: A single node (origin) with a message $M$ begins the process by sending $M$ to all its neighbours in the first round. These nodes will, in the second round, in parallel forward $M$ to all the other neighbours except the origin and so on. Nodes will do this forwarding in a mechanical manner not retaining any memory, thus, forwarding $M$ again if they receive it again. Possibly, the process can go on indefinitely.
We call this process \emph{Amnesiac Flooding (AF)} and define it more formally in later discussion.
How does Amnesiac flooding behave on the hypercube and Petersen graphs? What about other topologies?

Consider $AF$ on the hypercube first (Figure~\ref{fig: hcube}(c)) - it is easy to see that it stops after 3 rounds when the node diagonally opposite the origin gets $M$ from all of its neighbours simultaneously in round 3 and hence, cannot forward the message further. On the Petersen graph (Figure~\ref{fig: pgraph}(d)), though the process terminates, it takes 5 rounds and stops back at the origin. If we consider the termination times in terms of graph diameter, it takes diameter time on the hypergraph but much longer (2 times diameter plus 1) for the Petersen graph. 
Thus, the intriguing question: \emph{Will $AF$ terminate on other networks, and if so, how long will it take? Why does the time differ markedly on the Hypercube and the Petersen graphs though they are of similar sizes (in fact, the Petersen graph has a smaller diameter)?}


Flooding is among the most basic of distributed graph/network algorithms. To quote Apnes~\cite{AspnesFloodingWiki}:
\emph{Flooding is about the simplest of all distributed algorithms. It is dumb and expensive, but easy to implement, and gives you both a broadcast mechanism and a way to build rooted spanning trees.} At a high level, flooding can be simply described as:  \emph{In a network, a node begins flooding by sending a message to all its neighbours and subsequently, every node, in parallel, forwards the same message to all their neighbours}. 

Flooding is the simplest strategy to achieve broadcast i.e. have a message reach every node in the network, in quick time. 
Often flooding is implemented with a flag that is set when the message is seen for the first time to ensure termination  (see e.g.~\cite{Attiya-WelchBook})
We are interested in the variant of flooding which does not explicitly use such a flag or  keep a record of having seen the message before. The node selectively sends the message only to the complement of its neighbours from whom it has just received the message and subsequently forgets about that activity. The process terminates if there is no node that forwards $M$ in a round (and, therefore, subsequent rounds). We call this amnesiac flooding ($AF$ for short) to account for the very short term memory of the node. We analyse this very simple and theoretically interesting deterministic process on graphs and derive a rather unexpected and surprising result. We show that synchronous $AF$ (i.e.  in the synchronous message passing model) terminates on every finite graph in time optimal $O(d)$ rounds, where $d$ is the diameter of the graph. We also show that, at least in one asynchronous model, an adversary can force $AF$ to be non-terminating.

Besides being theoretically interesting, our results also have practical implications. 
$AF$ is a natural variant minimising memory overhead with nodes simply forwarding messages in a rather dumb manner.  Note that if there were multiple messages being flooded in the network, the memory requirement of keeping the historical flags could be significant, especially for low memory devices (e.g. sensor networks). Our results show that if the objective of the flooding is broadcasting, this overhead maybe unnecessary. Of course, a spanning substructure could be constructed from the initial regular flooding and used for subsequent broadcast (as is often done). However, spanning substructures can be difficult to maintain if the network is changing. This would not be required if $AF$ was being used for communication.

We speculate (though we have not studied this in detail) that amnesiac flooding may correspond to certain natural and social phenomena to whose understanding our results may contribute. Consider the following possibly contrived example as a thought experiment:
There is  an aggressive social media 
user that  forwards every message it receives to all its contacts but is polite enough to not forward  to those  who had just forwarded it the message. Naturally, such users lose track of the messages they have been forwarding.  A natural question is that will a message seize getting circulated.
These links  need to be investigated further.


\subsection{Model, Problem Definition and Results}  Let $G(V,E)$ be an undirected graph (with $n$ vertices and $m$ edges) representing a network where the vertices represent the nodes of the network and edges represent the connections between the nodes. 
 We consider the process in a synchronous message passing network: computation proceeds in synchronous rounds where each round consists of every node receiving messages from all its neighbours, doing local computation and sending messages to all (or some of) its neighbours. 
 No messages are lost in transit.  We consider only flooding from a single source for now.

\begin{definition}{\textbf{Synchronous Amnesiac Flooding (Synchronous AF):}}
A  distinguished node, say $\ell$, sends a message (say, $M$) to all its neighbours in round 1.  In subsequent rounds, every node receiving $M$ forwards a copy of $M$ to every, and only those, nodes it did not receive the message from in that round. Algorithm~\ref{algo:syncflood} presents the algorithm formally.

\begin{algorithm}
\begin{algorithmic}[1]
\STATE \textbf{procedure} $Flooding(G,s)$ \COMMENT{Flooding over graph G from source node s} 
\STATE Let $N(v) \leftarrow$ Neighbours of $v \in G$
\STATE Node $s$ sends message $M$ to all its neighbours in $G$ \COMMENT{Round 1: $s$ `floods' a message $M$}
\FOR{Rounds $i = 1, 2, \ldots$}
\FOR{For all nodes $v$ in parallel}
\STATE Let $I(v,M) \leftarrow$ set of neighbours of $v$  that sent $M$ to $v$ in round $i-1$ \COMMENT{$I(v,M) \subseteq N(v)$}
\STATE Send $M$ to $N(v) \setminus I(v,M)$ \COMMENT{Send to all neighbours except those who sent the message to $v$ in the previous round}
\ENDFOR
\ENDFOR
\end{algorithmic}
\caption{Synchronous Amnesiac Flooding: A message $M$ from a source node $s$ is `flooded' over graph $G$}
\label{algo:syncflood}
\end{algorithm}

\end{definition}

 Note that this is an `amnesiac' process i.e. nodes do not retain memory of having received or sent the message in the previous (but one) rounds.  
 We say that flooding \emph{terminates} when no message (i.e. a copy of $M$) is being sent over any edge in the network. We address the following questions:

\textbf{For every finite graph $G$, beginning from any arbitrary vertex, will amnesiac flooding always terminate? If so, how many rounds does it take?}
 
 
 
In Section~\ref{sec: synchterm},  we answer the first part of the above question in the affirmative   i.e. this flooding process will terminate for every $G$. For the second part of the question, in Section~\ref{sec: termtime}, we notice a sharp distinction between bipartite and non-bipartite graphs. We show that flooding terminates in $e$ rounds (i.e. at most $d$ rounds), where $e$ is the eccentricity of the source node and $d$ the diameter of $G$, if and only if $G$ is bipartite. Note that this is time optimal for broadcast. If the graph is non-bipartite, synchronous flooding takes longer: from a single source, flooding terminates in $j$ rounds where $e < j \leq e+d+1$. 
  in $D$ rounds on a bipartite graph
 
 Note that in this work, we only look at global termination i.e. the state when $M$ stops circulating in the system. We do not discuss the related problem of individual nodes detecting that either global termination has happened or if they should stop participation in flooding. In some sense, this is even unnecessary since nodes do not need to maintain any additional state or history. There is no persistent overhead to keeping the simple amnesiac flooding process as a rule in the background. 
 

\subsubsection{Asynchronous Message Passing}

For comparison, we also consider an asynchronous message passing model and show in Section~\ref{sec: asynchterm} that an \emph{adaptive} adversary in this model  can cause flooding to be non-terminating. We consider what we call as the \emph{round-asynchronous model} where the computation still proceeds in global synchronous rounds but the adversary can decide the delay of message delivery on any link.
The message cannot be lost and will be eventually delivered but the adversary can decide which round to deliver the message in. The adaptive adversary can decide on individual link delays for a round based on the state of the network for the present and previous rounds (i.e. node states, messages in transit and message history).  Now, the flooding algorithm (\emph{Asynchronous Amnesiac Flooding}) will exactly be same as Algorithm~\ref{algo:syncflood} except that the adversary decides which round a message transmitted on an edge reaches the other end. In  Section~\ref{sec: asynchterm}, we show that the adversary can force Asynchronous AF to be non-terminating by adaptively choosing link delays.

 We leave discussion of other asynchronous settings for future work.

\subsection{Some illustrative examples:}

\begin{figure}[h!]%
\centering 
\subfigure[Round 1]{ \includegraphics[scale=0.4]{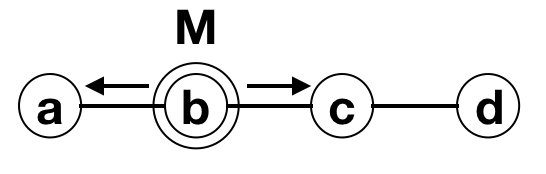}}\
\quad 
\subfigure[Round 2]{\includegraphics[scale=0.4]{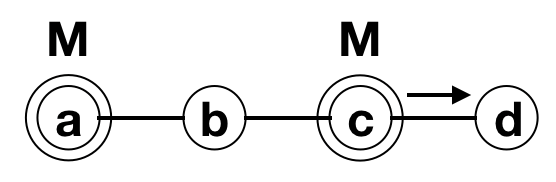}}
\quad
 \subfigure[Round 3]{  \includegraphics[scale=0.4]{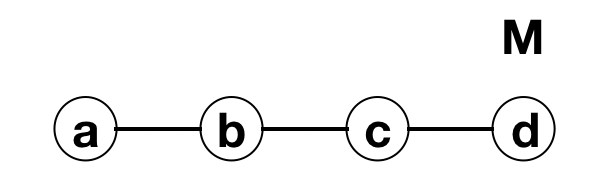}}
\caption{Amnesiac Flooding over a line network beginning with node $b$ in 2 ( $<$ diameter = 3)  rounds. Circled nodes are sending $M$ in that round.}
        \label{fig: floodingline}
\end{figure}

\begin{figure}[h!]%
\centering 
\subfigure[Round 1]{ \includegraphics[scale=0.3]{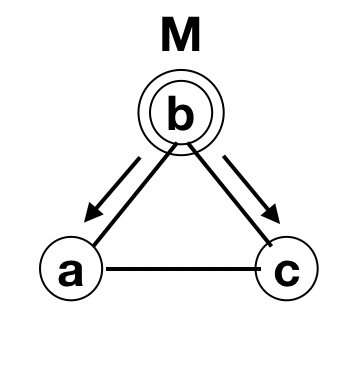}}\
\quad 
\subfigure[Round 2]{\includegraphics[scale=0.3]{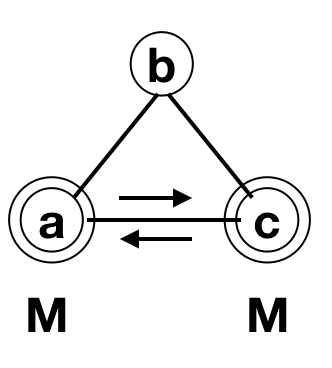}}
\quad
 \subfigure[Round 3]{  \includegraphics[scale=0.3]{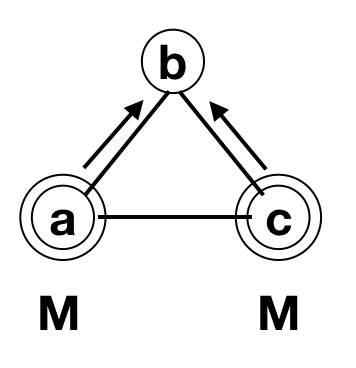}}
 \quad
  \subfigure[Round 4]{  \includegraphics[scale=0.3]{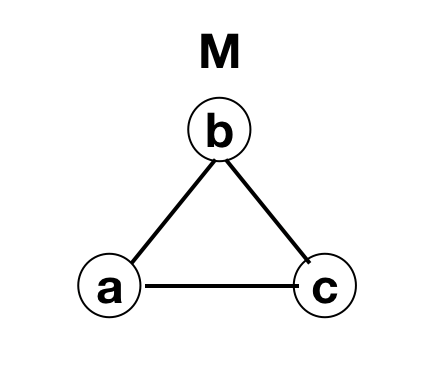}}
\caption{$AF$ over a Triangle (Odd Cycle/Clique) network beginning with node $b$. Both node $a$ and $c$ send $M$ to each other in round $2$ and to $b$ in round $3$. Also, this is an odd (\# nodes) cycle and termination takes $2d + 1$ time (d= diameter = 1). }
        \label{fig: floodingtriangle}
\end{figure}

Figure~\ref{fig: floodingline} shows flooding over a line graph. The process begins with the node $b$ and terminates at the ends of the graph and takes only 2 rounds, which is equal to the eccentricity of node $b$ in the graph (which has diameter of 3). Note that a line is an example of a bipartite graph. The triangle graph is another interesting illustrative example (Figure~\ref{fig: floodingtriangle}) -- here, termination takes 3 rounds, whereas, the diameter is only 1. Note that the triangle is also the smallest clique and the smallest non-trivial cycle with odd number of nodes (an important topology for us). The even cycle is another interesting topology but here termination will happen in $d$ rounds (as expected according to our bipartite graphs result). Of course, a graph can have far more complicated topology with cyclic and acyclic subgraphs. 

%


\subsection{Related work:}
The applications of flooding as a distributed algorithm are too numerous to be mentioned. It is one of the first algorithms to be introduced in distributed computing textbooks, often as the basic algorithm to solve leader election and set up graph substructures such as spanning trees~\cite{Lynchbook, Attiya-WelchBook, peleg, GerardTelDistributedAlgosBook}. Flooding based algorithms (or flooding protocols) appear in areas ranging rom GPUs, High performance, shared memory and parallel computing to  Mobile ad hoc networks(MANETs), Mesh Networks, Complex Networks etc~\cite{tanenbaum2011computer}. In~\cite{Rahman04controlledflooding}, Rahman et al show that  flooding can even be adopted as a reliable and efficient routing scheme, comparable to  sophisticated point-to-point forwarding schemes, in some ad-hoc wireless mobile network applications.

Since it is imperative to not have unnecessary messages circulating and clogging the network, explicit termination is desired and often enforced by using a flag to record if the node has already participated in the flooding ~\cite{AspnesFloodingWiki, Lynchbook, Attiya-WelchBook, peleg,GerardTelDistributedAlgosBook}. Variants of flooding Termination is one of the most important properties a distributed algorithm requires. However, in some models such as population protocols, the low memory makes termination very difficult to achieve leading to research that tries to provide termination e.g.~\cite{Michail-SpirakisPP2015}. Our flooding algorithm has the advantage of being simple, using low memory, and being efficiently terminating as shown by our analysis. The idea of avoiding the most recently chosen node(s) has been used before in distributed protocols e.g. in social networks~\cite{Doerr-Social-EuroComb11} and broadcasting~\cite{Elsasser-Memory-SODA08}
but we are not aware of this fundamental variant of flooding having been studied before.



%

\section{Termination in a synchronous network} 

\label{sec: synchterm}

\begin{definition}
\label{def: roundsets}
Let $G$ be the graph. The {\it round-sets} $R_0, R_1, \ldots $ are defined as:
\[
\begin{array}{lll}
R_0 & \hbox{\it is the singleton containing the initial node,} & \; \\
R_i & \hbox{\it is the set of nodes which receive a message at round i} & (i\geq 1). \\
\end{array}
\]
\end{definition}Clearly, if $R_j = \emptyset$ for some $j \geq 0$, then $R_i = \emptyset$ for all $i \geq j$. We shall refer to rounds $R_i$, where $R_i \neq \emptyset$, as {\it active} rounds.

\begin{theorem}
\label{th: twodistinct}
Any node $g \in G$ is contained in at most two distinct round-sets.
\end{theorem}

\begin{proof}
Define ${\cal R}$ to be the set of finite sequences of consecutive round-sets of the form:

\begin{equation}
\label{eq: roundset}
\underline{R} = 
R_s, \ldots, R_{s+d} \;\;\; \hbox{ {\it where} $s\geq 0, \;d>0$, {\it and} $R_s \cap R_{s+d} \neq \emptyset$ }.
\end{equation}

In~(\ref{eq: roundset}), $s$ is the {\it start-point} $s(\underline{R})$ and $d$ is the {\it duration} $d(\underline{R})$ of $\underline{R}$. Note that, a node $g \in G$ belonging to $R_s$ and $R_{s+d}$  may also belong to other $R_i$ in ~(\ref{eq: roundset}).  If a node $g \in G$ occurs in three different round-sets $R_{i_1}$, $R_{i_2}$ and $R_{i_3}$, then the duration between $R_{i_1}$ and $R_{i_2}$, $R_{i_2}$ and $R_{i_3}$, or $R_{i_1}$ and $R_{i_3}$ will be even. Consider the subset ${\cal R}^e$ of ${\cal R}$ of sequences of the form~(\ref{eq: roundset}) where $d$ is even. To prove that no node is in three round-sets, it suffices to prove that ${\cal R}^e$ is empty. 

We assume that ${\cal R}^e$ is non-empty and derive a contradiction. 

Let ${\cal R}^e_{md}$ be the subset of ${\cal R}^e$ comprising sequences of minimum (even) duration $md$, i.e.
\begin{equation}
{\cal R}_{md}^e = \{ \underline{R} \in {\cal R}^e \;\; | \;\; \forall \; \underline{R}'\in {\cal R}^e . \;\; d(\underline{R}') \geq d(\underline{R})=md \}
\end{equation}
Clearly, if ${\cal R}^e$ is non-empty then so is ${\cal R}^e_{md}$. Let $\underline{R}^* \in {\cal R}^e_{md}$ be the sequence with earliest start-point $ms$,  i.e.
\begin{equation}
\underline{R}^* = R_{ms}, \ldots , R_{ms+md}
\end{equation}where
\begin{equation}
 \forall \; \underline{R}'\in {\cal R}_{md}^e\;.\; s(\underline{R}') \geq s(\underline{R}^*) = ms 
\end{equation}

By~(\ref{eq: roundset}), there exists $g \in R_{ms} \cap R_{ms+md}$. Choose node $g'$ which sends a message to $g$ in round $ms+md$. As $g'$ is a neighbour of $g$, either $g'$ sends a message to $g$ in round $ms$ or $g$ sends a message to $g'$ in round $ms+1$. We show that each of these cases leads to a contradiction.


\begin{itemize}
\item[Case (i):] $g'$ sends a message to $g$ in round $ms$ (Figure~\ref{fig: case1})

\begin{figure}[h!]
\centering
\includegraphics[scale=0.5]{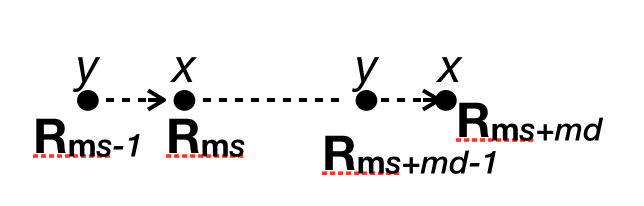}
\caption
{Node $g'$ sends a message to node $g$ in round $ms$: the first round of the minimum even length sequence (of length $md$) in which $g$ repeats}
\label{fig: case1}
\end{figure}

\noindent In this case,  there must be a round $ms-1$ which is either round 0 and $g'$ is the initial node, or $g'$ received a message in round $ms-1$.  Thus, the sequence 
\begin{equation}
\underline{R}^{*'} = R_{ms-1},R_{ms},\ldots, R_{ms+md-1}\;\;\; \hbox{ {\it where}}\; g' \in R_{ms-1} \cap R_{ms+md-1} 
\end{equation}has $d( \underline{R}^{*'})=(ms+md-1)-(ms-1) =md$ which is even and so $\underline{R}^{*'} \in {\cal R}_{md}^e$. As $\underline{R}^{*'} \in {\cal R}_{md}^e$,  by (4)
\begin{equation}
s(\underline{R}^{*'}) \geq s(\underline{R}^*)
\end{equation}But, from (5), $s(\underline{R}^{*'}) = ms-1$ and, from (4), $ s(\underline{R}^*)=ms$. Thus, by (6),
\[
ms-1 = s(\underline{R}^{*'}) \geq s(\underline{R}^*) = ms
\]
which is a contradiction.

$\;$

\item[Case (ii):] $g$ sends a message to $g'$ in round $ms+1$ (Figure~\ref{fig: case2})

\begin{figure}[h!]
\centering
\includegraphics[scale=0.5]{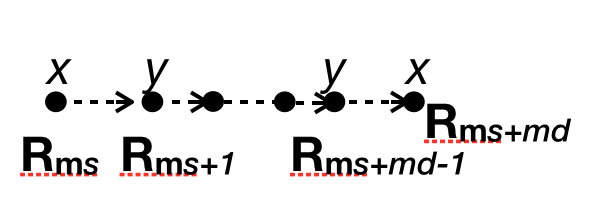}
\caption
{Node $g$ sends a message to node $g'$ in round $ms+1$: round $ms$ is the first round of the minimum even length sequence (of length $md$) in which $g$ repeats}
\label{fig: case2}
\end{figure}

\noindent  By the definition of ${\cal R}^e$, the smallest possible value of $md$ is 2. However, it is not possible to have $md=2$ in this case as then
\[
\underline{R}^* = R_{ms}, R_{ms+1}, R_{ms+2}
\]and $g$ sends a message to $g'$ in round $ms+1$ and we chose $g'$ to be such that $g'$ sends a message to $g$ in round $ms+md=ms+2$,  which cannot happen as $g$ cannot send a message to $g'$ and $g'$ to $g$ in consecutive rounds by the definition of rounds. So,
\[
\underline{R}^* = R_{ms}, R_{ms+1},\ldots, R_{ms+md-1}, R_{ms+md}
\] where $ms+1<ms+md-1$. Consider the sequence
\begin{equation}
\underline{R}^{*''} =  R_{ms+1},\ldots, R_{ms+md-1}
\end{equation}As $g'$ receives a message from $g$ in round $ms+1$ and $g'$ sends a message to $g$ in round $ms+md$, it is clear that $g' \in 
 R_{ms+1} \cap R_{ms+md-1}$. Thus, $\underline{R}^{*''} \in {\cal R}$. As $md$ is even, so is $(ms+md-1)-(ms+1)=md-2$ and therefore  $\underline{R}^{*''} \in {\cal R}^e$. Now, $\underline{R}^* \in {\cal R}^e_{md}$ and so, as $\underline{R}^{*''} \in {\cal R}^e$, we have,  by (2), 
\begin{equation}
 d(\underline{R}^{*''}) \geq d(\underline{R}^*)
\end{equation}As $d(\underline{R}^{*''})=md-2$ from (7) and $ d(\underline{R}^*) = md$ from (3), we have, by (8),
\[
md-2 =  d(\underline{R}^{*''}) \geq d(\underline{R}^*) = md
\]

This contradiction completes the proof.

\end{itemize}

\end{proof}

\begin{definition}
Given $g \in G$, we use a superscript 1 to indicate that $g$ belongs to a round-set for the first time, and a superscript 2 to indicate that it belongs to a round-set for the second time, i.e.\[
g^1 \in R_j
\]means that 
\[
g \in R_j \;\;\; {\it and} \;\;\;g \notin R_i \;\;\; \hbox{{\it for all i with}} \;\; 0 \leq i < j.
\]and
\[
g^2 \in R_j
\]means that 
\[
g \in R_j \;\;\; {\it and} \;\;\;g \in R_i \;\;\; \hbox{{\it for some (unique by Theorem~\ref{th: twodistinct}) i with}} \;\; 0 \leq i < j.
\]
\end{definition}

Theorem~\ref{th: twodistinct} implies that $R_i = \emptyset$ for $i \geq 2|G|$, where $|G|$ is the order (number of vertices) of $G$, and therefore network flooding always terminates.

\begin{corollary}
Synchronous network flooding always terminates in fewer than $2|G|+1$ rounds.
\end{corollary}In the next section we give a greatly improved sharp upper bound for the number of rounds to termination, in terms of the eccentricity of the initial node and the diameter of $G$.

\section{Time to termination}
\label{sec: termtime}

\begin{figure}[h!]%
\centering 
\subfigure[Even Cycle Graph]{ \includegraphics[scale=0.5]{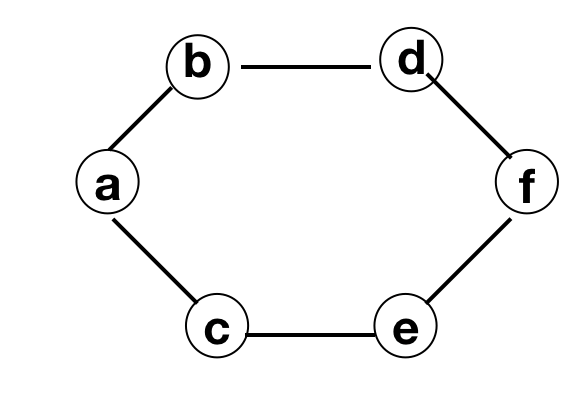}}
\label{fig: even}
\qquad 
\subfigure[Odd Cycle Graph]{\includegraphics[scale=0.5]{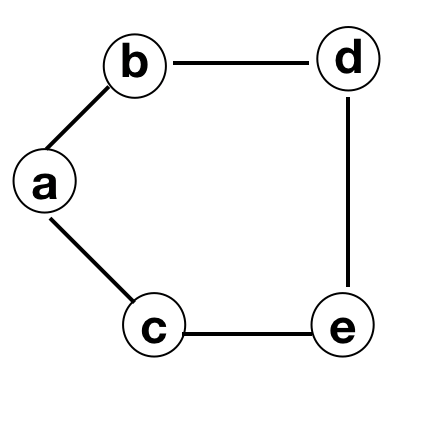}}
\label{fig: odd}
\caption{Even cycle (6 nodes) and Odd cycle (5 nodes) graphs: Graphs show markedly different termination times. Consider $AF$ from node $b$ in both cases - In the 6-cycle it terminates in 3 rounds, but in the 5-cycle in 5 rounds.}
        \label{fig: Odd-EvenCycles}
\end{figure}

The question of termination of network flooding is non-trivial when cycles are present in $G$. The simple cases when $G$ is an even cycle, as in Figure~\ref{fig: even}(a)
and when $G$ is an odd cycle, as in Figure~\ref{fig: odd}(b)
 display quite different termination behaviours. The even cycle in Figure~\ref{fig: even}(a) terminates remotely from an initial node, after round $e$ where $e$ is the eccentricity of the initial node in $G$. On the other hand, flooding on the odd cycle in Figure~\ref{fig: odd}(b), returns a message to the initial node and terminates after round $2e+1$ resulting in a longer flooding process than the even cycle in Figure~\ref{fig: even}(a) despite having fewer nodes and a smaller value of $e$. In this section, we show that these observations can be largely generalized to arbitrary graphs. Specifically, we show that flooding on a graph $G$ terminates after $e$ rounds if and only if $G$ is bipartite. If $G$ is not bipartite, we show that flooding terminates after some round $i$ where $e < i \leq e+d+1$ and $d$ is the diameter of $G$. 
 
\begin{definition}
\label{def: ecdistances}
Let $(G,E)$ be a graph with vertex set G and edge set E, and $g_0 \in G$ be an initial node. We will use the following definitions.
\begin{itemize}
\item[(i)]
\label{def: ecdistances-1}
For each $j \in \mathbb{N}$, the distance set $D_j$ will denote the set of points which are a distance $j$ from $g_0$. i.e.
\[
D_j = \{ g \in G \;   : \; d(g_0,g) = j \},
\]where $d$ is the usual distance function in graph $G$.
\item[(ii)]
\label{def: ecdistances-2}
A node $g \in G $ is an equidistantly-connected node, abbreviated ec node, iff there there exists $g' \in G- \{ g_0,g \}$ such that $d(g_0,g)=d(g_0,g')$ and $\{ g, g' \} \in E$
\end{itemize}
\end{definition}

We have the following basic properties of distance sets $D_j$ and $ec$ nodes.

\begin{lemma} Let $G$ be a graph and $g_0 \in G$ be an initial node.
\label{lm: eclemma}
\begin{itemize}
\item[(i)]
For all $j  \in \mathbb{N}$ and $i>j$, $D_j \subseteq R_j$ and $R_j \cap D_i = \emptyset$. \label{lm: eclemma-1}
\item[(ii)]
For all $j  \in \mathbb{N}$, $g \in D_j$ and $g' \in D_{j+1}$ such that $g$ and $g'$ are neighbours, $g$ sends a message to $g'$ in round $j+1$, i.e. all nodes at a distance $j$ from $g_0$ send to all their neighbours which are a distance $j+1$ in round $j+1$. \label{lm: eclemma-2}
\item[(iii)]
\label{lm: eclemma-3}
If $j \geq 1$ and $g \in D_j$ is an $ec$ point, then $g^2 \in R_{j+1}$.
\end{itemize}
\end{lemma}

\begin{proof}
For (i), we prove the following assertion by induction:
\begin{equation}
\label{eq: round-Ind}
\hbox{{\it for all $j  \in \mathbb{N}$, $j' \leq j$, $i>j$, $D_j \subseteq R_j$ and $R_{j'} \cap D_i = \emptyset$}}.
\end{equation}

\begin{itemize}
\item {\it Case $j=0$:}

Firstly, if $j=0$, then $D_j = D_0 = \{g_0 \} = R_0 = R_j$ . Also, for all $j' \leq j =0$ and $i > j = 0$, $R_{j'} \cap D_i = R_0 \cap D_i = \emptyset$ as $R_0 = \{ g_0 \}$ and $D_i$ is the set of nodes a distance $i>0$ from $g_0$ (Definition~\ref{def: ecdistances-1}).

\noindent {\it Inductive step}

\noindent Assume \ref{eq: round-Ind} 
  holds for some case $j \geq0$. We show that implies that (9) holds for the case $j+1$, i.e.
\begin{equation}
\hbox{{\it for all $j  \in \mathbb{N}$, $j' \leq j+1$, $i>j+1$, $D_j \subseteq R_j$ and $R_{j'} \cap D_i = \emptyset$}}.
\end{equation} First of all, we prove that 
\begin{equation}
\hbox{{\it for all $j  \in \mathbb{N}$, $j' \leq j+1$, $i>j+1$,  $R_{j'} \cap D_i = \emptyset$}}
\end{equation}By induction, we already have that
\begin{equation}
\hbox{{\it for all $j  \in \mathbb{N}$, $j' \leq j$, $i>j$, $R_{j'} \cap D_i = \emptyset$}}
\end{equation}All that is needed to prove (11) from (12) is to show that 
\begin{equation}
\hbox{$R_{j+1} \cap D_i = \emptyset$ {\it for all} $i> j+1$}
\end{equation}Assume, on the contrary, that there is a 
\begin{equation}
\hbox{$g \in R_{j+1} \cap D_i $ {\it for some} $i>j+1$}
\end{equation}Then, $g$ receives a message from a neighbour $g'$ in round $j+1$. So, $g'$  received a message in round $j$, i.e. $g' \in R_j$. As $g \in D_i$ is a distance $i$ from $g_0$, its neighbour $g'$ must belong to one of $D_{i-1}$, $D_i$ or $D_{i+1}$. Thus, 
\begin{equation}
\hbox{$g' \in R_j \cap D_{i-1}$ {\it or} $g' \in R_j \cap D_{i}$ {\it or} $g' \in R_j \cap D_{i+1}$}
\end{equation}All three possibilities in (15) contradict the inductive hypothesis at (12)  (the first case in (15) contradicts (12) as $i>j+1$, from (14), and so $i-1 > j$) and so (13) and therefore (11) holds. We now prove
\begin{equation}
\hbox{{\it for all $j  \in \mathbb{N}$, $j' \leq j+1$,  $D_j \subseteq R_j$}}.
\end{equation}

which, along with (11) (which we have just proved), will establish the whole statement of the inductive step at (10). By induction, we already have that

\begin{equation}
\hbox{{\it for all $j  \in \mathbb{N}$, $j' \leq j$,  $D_j \subseteq R_j$}} \label{eq: Dsubset}
\end{equation}

All that is needed to prove (16) from (17) is to show that 

\begin{equation}
\hbox{$D_{j+1} \subseteq R_{j+1}$}
\end{equation}

\item {\it Case $j=1$:}

 By \ref{eq: Dsubset},  
  $D_1 \subseteq R_1$ from round $1$ and the nodes in $D_1$ only receive messages from the initial node. In particular, no messages are received from nodes in $D_2$. Hence, in round 2, the nodes in $D_1$ send messages to all the nodes in $D_2$ which are all neighbours of nodes in $D_1$ by Definition~\ref{def: ecdistances-1}. Thus, $D_2 \subseteq R_2$. If $j>1$, by (17), $D_j \subseteq R_j$ from round $j$ and nodes in $D_j$ do not receive messages from any neighbours in $D_{j+1}$  as, by (12), $R_{j-1} \cap D_{j+1} = \emptyset$  and so nodes in $D_{j+1}$ do not have messages in round $j-1$ to send in round $j$. Hence, in round $j+1$, the nodes in $D_j$ send messages to all their neighbours  in $D_{j+1}$ and so $D_{j+1} \subseteq R_{j+1}$ as required by (18), and the proof of the inductive step (10) is complete.

\end{itemize}

For (ii), we note that the only circumstance in which a node $g$ in $D_j$ ($\subseteq R_j$ by (i)) does not send to a neighbour $g'$ in $D_{j+1}$ in round $j+1$ is if $g$ sent a message to $g'$ in round $j$. This would need $g$ to be in the round-set $R_{j-1}$, i.e. $g \in R_{j-1} \cap D_{j+1}$ which contradicts (i) which has $R_{j-1} \cap D_{j+1} = \emptyset$ as $j+1 > j-1$.
 
For (iii), if $j \geq 1$ and $g \in D_j$ is an $ec$ point, then by Definition 2.2(i) there is an point $g'$ equidistant from the initial node $g_0$, i.e. $g' \in D_j$ such that $g$ and $g'$ are neighbours. By (i) of this lemma $D_j \subseteq R_j$, and so both $g$ and $g'$ receive messages in round $j$.  Also by (i), neither sends a message in round $j$ as $R_{j-1} \cap D_j = \emptyset$. Thus, $g$ and $g'$ send messages to each other in round $j+1$. As this will be the second time they receive messages we have that $g^2 \in R_{j+1}$.

\end{proof}

$\;$

\noindent All nodes in a graph without $ec$ nodes, belong to at most one round-set.

\begin{lemma}
\label{lm: noec}
If $G$ is a graph, then $G$ has no $ec$ nodes if and only if each node in $G$ is contained in exactly one round-set.
\end{lemma}

\begin{proof}
Suppose that $G$ has no $ec$ nodes. Assume, on the contrary, that $G$ has nodes that appear in two round-sets.  Let $R_j$ $(j \geq 1)$ be the earliest round which contains a node $g$ such that $g^2 \in R_j$ and $h \in R_{j-1}$ be a neighbour of $g$ which sends to $g$ in round $j$, so that $h^1 \in R_{j-1}$. Then, $h \in D_i$ for some $i \geq 1$ and $h^1 \in R_i$ by Lemma~\ref{lm: eclemma-1}.  Thus, $i=j-1$ and so $g^2 \in R_{i+1}$. As $g$ is a neighbour of $h$, $g \in D_i$, $D_{i+1}$, or $D_{i-1}$. If $g \in D_i$ then $g$ and $h$ are $ec$ nodes contrary to our supposition that $G$ has no $ec$ nodes. If $g \in D_{i+1}$ then $g^1 \in R_{i+1}$ by Lemma~\ref{lm: eclemma-1}, which is contrary to the assertion that $g^2 \in R_j = R_{i+1}$. If $g \in D_{i-1}$ then $g \in R_{i-1}$, by Lemma~\ref{lm: eclemma-1}, and so $g^1 \in R_{i-1}$ as $g^2 \in R_{i+1}$. By Lemma~\ref{lm: eclemma-2}, $g$ sends to $h$ in round $i=j-1$. This is contrary to $h$ sending to $g$ in round $j$. Thus, our assumption that $G$ has nodes that appear in two round-sets is false.

Conversely, suppose that $G$ has an $ec$ node $g$, $g \in D_j$ say where $j \geq 1$. Then $g^2 \in R_{j+1}$ by Lemma~\ref{lm: eclemma-3}.
\end{proof}

$\;$

\noindent We note that bipartite graphs do not have any $ec$ nodes.
\begin{lemma}
\label{lm: ecbipartite}
Let $G$ be a graph  and $g_0 \in G$ be an initial node. Then, $G$ is bipartite iff it has no $ec$ nodes.
\end{lemma}
\begin{proof}
It is easy to see that nodes equidistant from the initial node must belong to the same partite set. A graph is bipartite iff no edge connects two such nodes, and this is the case iff $G$ has no $ec$ nodes by Definition~\ref{def: ecdistances-2}.
\end{proof}

$\;$

\noindent From Lemmas~\ref{lm: noec} and~\ref{lm: ecbipartite}, we see that, in bipartite graphs, nodes only appear in one round-set. Thus, the time to termination can be determined by finding a bound on when each node belongs to a round-set.

\begin{theorem}
\label{th: bipartitetermtime}
Let $G$ be a graph and $g_0 \in G$ be an initial node with eccentricity $e$. Then, flooding will have terminated after round $e$ if and only if $G$ is bipartite.
\end{theorem}

\begin{proof}
\[
\begin{array}{llll}
\hbox{$G$ is bipartite} & \hbox{iff} & \hbox{$G$ has no $ec$ nodes} & \hbox{(by Lemma~\ref{lm: ecbipartite})} \\
\; &\hbox{iff} & \hbox{no node appears in 2 round-sets} &  \hbox{(by Lemma~\ref{lm: noec})}   \\
\; & \hbox{iff} & \hbox{$R_e$ is the last non-empty round-set} &  \hbox{(by Lemma~\ref{lm: eclemma-1})}  \\
\end{array}
\]
\end{proof}

\noindent To find the time to termination in general graphs we need to find a bound on when nodes can belong to a round-set for the second time. As nodes can only belong to at most two round-sets, by Theorem~\ref{th: twodistinct}, this will give a bound for termination of flooding in general graphs. The following lemma relates the round-sets of second occurrences of neighbouring nodes.
\begin{lemma} 
\label{lm: gsquare}
Let $G$ be a graph and $g_0 \in G$ an initial node. If $h \in G$ and $h^2 \in R_j$ for some $j  \in \mathbb{N}$, and if $g$ is a neighbour of $h$, then
\[
g^2 \in R_{j-1} \;\; {\it or} \;\; g^2 \in R_{j}  \;\; {\it or} \;\; g^2 \in R_{j+1}
\]
\end{lemma}
\begin{proof}
Let $i$ be the distance of $h$ from $g_0$, i.e. $h \in D_i$. Then, as $h^2 \in R_j$, $j>i$ by Lemma~\ref{lm: eclemma-1}. As $g$ is a neighbour of $h$, $g \in D_{i}$ or $g \in D_{i-1}$ or $g \in D_{i+1}$.

\begin{itemize}
	\item {\it Case $g \in D_i$:} As $h,g \in D_i$ are neighbours they are both $ec$ nodes. Thus, by Lemma~\label{lm: eclemma-3}, $h^2 \in R_{i+1}$ and $g^2 \in R_{i+1}$. Therefore, $j = i+1$ and $g^2 \in R_j$.

\item {\it Case $g \in D_{i-1}$:} If $g \in R_j$ ($\neq R_{i-1}\;\hbox{as} \; j>i$) then, as $g^1 \in D_{i-1} \subseteq R_{i-1}$ by Lemma~\ref{lm: eclemma-1}, it must be the case that $g^2 \in R_j$. If $g\notin R_j$ and $g \in R_{j-1}$ ($\neq R_{i-1}\;\hbox{as} \; j>i$) then, as $g^1 \in   R_{i-1}$ by Lemma~\ref{lm: eclemma-1}, it must be the case that $g^2 \in R_{j-1}$. If $g\notin R_j$ and $g \notin R_{j-1}$ then, as $h \in R_j$, $h$ sends to $g$ in round $j+1$ and so $g \in R_{j+1}$ ($\neq R_{i-1}\;\hbox{as} \; j>i$). As $g^1 \in R_{i-1}$, it must be the case that $g^2 \in R_{j+1}$.

\item {\it Case $g \in D_{i+1}$, $g$ does not send to $h$ in round $j$:}
 In this case, as $h \in R_j$, $h$ sends to $g$ in round $j+1$. Thus, $g \in R_{j+1}$ ($\neq R_{i+1}\;\hbox{as} \; j>i$) and therefore, as $g^1 \in D_{i+1} \subseteq R_{i+1}$ by Lemma~\ref{lm: eclemma-1}, it must be the case that $g^2 \in R_{j+1}$.

\item {\it Case $g \in D_{i+1}$, $g$ sends to $h$ in round $j$:}
In this case $g \in R_{j-1}$. We show that $g^1 \notin R_{j-1}$. Assume, on the contrary, that $g^1 \in R_{j-1}$. Then, by Lemma~\ref{lm: eclemma-1}
, $g^1 \in D_{i+1} \subseteq R_{i+1}$ and thus $j-1 = i+1$. Hence, by Lemma~\ref{lm: eclemma-1}, $h^1 \in D_i \subseteq R_i = R_{j-2}$. Also, $g \notin R_{j-3}$ as $g^1 \in R_{j-1}$. 

To summarize:
\[
g \notin R_{j-3},\;\; h^1 \in R_{j-2},\;\; g^1 \in R_{j-1},\;\; h^2 \in R_j
\]So, $h$ sends to $g$ in round $j-1$ and $g$ sends to $h$ in round $j$ by the case assumption. This is a contradiction. Thus, the assumption that $g^1 \in R_{j-1}$ is false and, as $g\in R_{j-1}$, it follows that $g^2 \in R_{j-1}$.
\end{itemize}
 This completes the proof.
\end{proof}

\begin{theorem}
\label{th: nonbiterminationtime}
Let $G$ be a non-bipartite graph with diameter $d$ and let $g_0 \in G$ be an initial node of eccentricity $e$. Then, flooding terminates after $j$ rounds where $j$ is in the range $e < j \leq e+d+1$.
\end{theorem}

\begin{proof}
If $G$ is not bipartite it has an $ec$ node $g$, by Lemma~\ref{lm: ecbipartite}.  By Lemma~\ref{lm: eclemma-3}, $g^2 \in R_k$ where $k = d(g_0,g) +1$. Let $h$ be an arbitrary node in $G$ other than $g$. Then, there is a path
\[
h_0=g \longrightarrow h_1 \longrightarrow \ldots \longrightarrow h_l = h
\]where $l \leq d$. By repeated use of Lemma~\ref{lm: gsquare}, 
\[
\begin{array}{llll}
h_1^2 \in R_{j_1} & \hbox{where} & k-1 \leq j_1 \leq k+1, & \; \\
h_2^2 \in R_{j_2} & \hbox{where} & j_1-1 \leq j_2 \leq j_1+1, & \; \\
\ldots & \; & \; & \; \\
h_l^2 \in R_{j_l} & \hbox{where} & j_{l-1}-1 \leq j_l \leq j_{l-1}+1 & (l \geq 1). \\
\end{array}
\]Thus, 
\begin{equation}
h^2_l \in R_{j_l} \;\;\; {\it where} \;\;\; k-l \leq j_l \leq k+l
\end{equation}Put $j = j_l$. From (19), as $k = d(g_o,g) +1 \leq e+1$ and as $l \leq d$,
\[
h^2_l \in R_{j} \;\;\; {\it where} \;\;\;j \leq e+d+1
\]As $G$ is not bipartite, $j>e$ by Theorem~\ref{th: bipartitetermtime} and the proof is complete.
\end{proof}

$\;$

\begin{figure}[th!]
\centering
\includegraphics[scale=0.5]{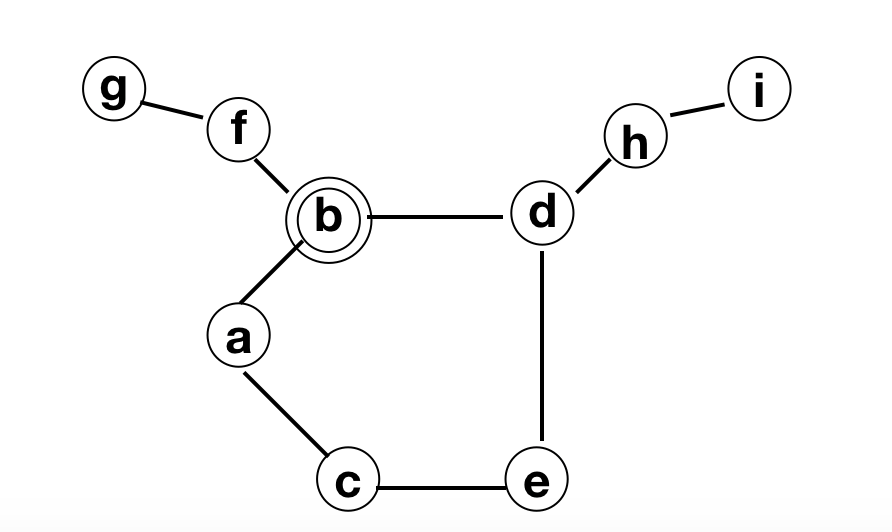}
\caption
{Flooding in the graph in the above figure starting from node $b$ takes $e + d + 1$ rounds (the maximum as per our analysis), where $e$ is eccentricity and $d$ the diameter.}
\label{fig: 5cyclecat}
\end{figure}

\noindent The upper bound in Theorem~\ref{th: nonbiterminationtime} is easily seen to be sharp - the flooding in the graph in Figure~\ref{fig: 5cyclecat} starting from node $b$ terminates after round 7 = $2+4+1 = e+d+1$. Similar termination times hold for all nodes in the Petersen graph (Figure~\ref{fig: pgraph}).

%
%
%
%

\section{Asynchronous Amnesiac Flooding}
\label{sec: asynchterm}

\begin{figure}[h!]%
\centering 
\subfigure[Round 1]{ \includegraphics[scale=0.3]{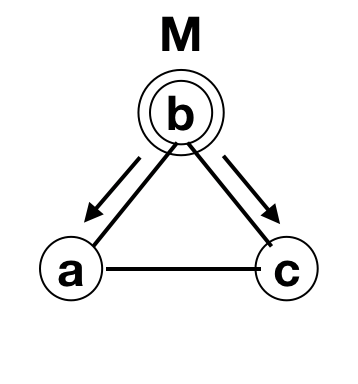}}\
\quad 
\subfigure[Round 2]{\includegraphics[scale=0.3]{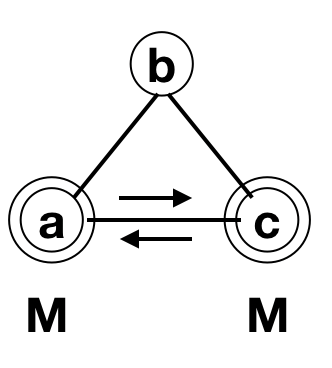}}
\quad
 \subfigure[Round 3]{  \includegraphics[scale=0.3]{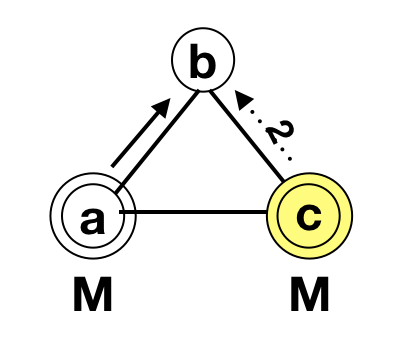}}
 \quad
  \subfigure[Round 4]{  \includegraphics[scale=0.3]{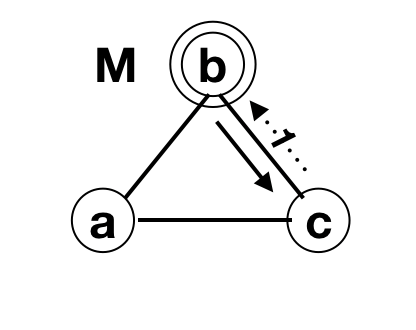}}
   \quad
  \subfigure[Round 5]{  \includegraphics[scale=0.3]{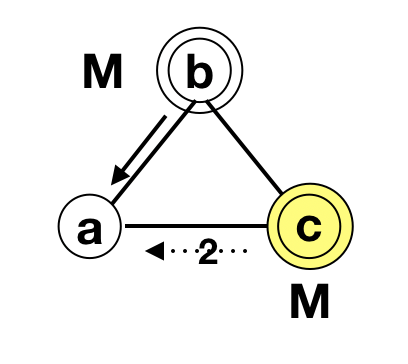}}
\caption{Asynchronous $AF$ over a Triangle. Both node $a$ and $c$ send $M$ to each other in round $2$. In round $3$,  $a$ sends $M$ to $b$ but the adversary makes $c$ holds the message for one round (shaded node). In the next round, we have a round analogous to round $2$ and so on.}
        \label{fig: floodingtriangle}
\end{figure}

\noindent \textbf{Non-termination in an adversarial asynchronous setting:} Consider the \emph{round-asynchronous} setting (as described in the model section). The scheduling adversary can adaptively choose the delay on every message edge i.e. which round to forward a message on. 

An example suffices to prove non-termination. Consider round 3 in the triangle in Figure~\ref{fig: floodingtriangle}. The adversary delays $M$ at node $c$ but $a$ continues and sends to $b$.  In round 4, node $b$ and $c$ both send $M$ so that the beginning of the next round is now identical to round $2$ with nodes $a$ and $b$ interchanged. This process can now continue \emph{ad infinitum} with the adversarial intervention.

\section{Conclusion and Future Work}


We studied a natural variant of the flooding algorithm where nodes do not retain any memory of the flooding beyond the previous round. We call this Amnesiac flooding ($AF$) and discussed the question of termination i.e. no copies of the initial message are being circulated anymore. We showed the surprising result that not only does this process  terminate on all finite graphs but also accomplishes broadcast in almost optimal time and message overhead. There is a clear separation in complexity between bipartite and non-bipartite topologies. An interesting question is whether this separation can be exploited to devise distributed procedures to detect the topology of a graph given distance measures or vice versa. We have not addressed the question of multiple sources: what happens when multiple nodes start the flooding process with the same message $M$? What about dynamic settings where nodes and edges change? It is easy to see that due to its simplicity, $AF$ can be re-executed immediately after the graph has changed. However, what if the graph changes while messages are in circulation  - under what conditions is termination/non-termination guaranteed?

Another important question is to look at flooding in asynchronous settings in more detail. We show one model where an adversary can force $AF$ to be non-terminating. Since a completely asynchronous setting is event driven, this would also involve deciding what it means to receive messages simultaneously. Finally, one can see processes such as random walks, coalescing random walks and diffusion as probabilistic extremal variants of flooding. Are there any implications  or connections of our result on these or intermediate probabilistic models? What about randomised variants of $AF$?

\section*{Acknowledgements} We would like to thank Saket Saurabh, Jonas Lefevre, Chhaya Trehan, Gary Bennett, Valerie King, Shay Kutten, Paul Spirakis, Abhinav Aggarwal for the useful discussions and insights and to all others in our network who attempted to solve this rather easy to state puzzle.

\bibliography{selfheal-routing}
\bibliographystyle{plain}

\end{document}